\documentclass[floatfix,twocolumn,prl,tightenlines,preprintnumbers,amsmath,amssymb,superscriptaddress]{revtex4-2} 
\usepackage{amsmath}
\usepackage{bm}
\usepackage{graphicx}
\usepackage{mathrsfs}
\usepackage{multirow}
\usepackage{graphicx}
\usepackage{booktabs, ctable}
\usepackage{xcolor, color, framed}
\usepackage[colorlinks, linkcolor=red,anchorcolor=green,citecolor=blue]{hyperref}

\begin{document}
\title{Color screening versus thermal decay as the mechanism of $\Upsilon$ suppression\\ in high energy nuclear collisions}
	
\author{Yida Yang}
\email{yidayang@tju.edu.cn}
\affiliation{Department of Physics, Tianjin University, Tianjin 300354, China}

\author{Baoyi Chen}
\email{baoyi.chen@tju.edu.cn}
\affiliation{Department of Physics, Tianjin University, Tianjin 300354, China}

\author{Jiaxing Zhao}
\email{jzhao@itp.uni-frankfurt.de}
\affiliation{Helmholtz Research Academy Hesse for FAIR (HFHF), GSI Helmholtz Center for Heavy Ion Physics, Campus Frankfurt, 60438 Frankfurt, Germany}
\affiliation{Institut f\"ur Theoretische Physik, Johann Wolfgang Goethe-Universität,Max-von-Laue-Straße 1, D-60438 Frankfurt am Main, Germany}

\author{Pengfei Zhuang}
\email{zhuangpf@mail.tsinghua.edu.cn}
\affiliation{Department of Physics, Yantai University, Yantai 264005, China}
\affiliation{Department of Physics, Tsinghua University, Beijing 100084, China}
\affiliation{South Center for Nuclear-Science Theory, Institute of Modern Physics, Chinese Academy of Sciences, Huizhou 516000, China}

\begin{abstract}
To clearly identify the mechanism behind the suppression of heavy quarkonium in relativistic heavy-ion collisions, we study $\Upsilon$ production at RHIC energies by solving its transport equation driven solely by the suppression rates. By calculating the nuclear modification factor and comparing it with experimental data, we find that the sudden suppression governed by the color-screening temperature cannot simultaneously describe both the ground and excited states of the $\Upsilon$, whereas the continuous suppression induced by thermal decay successfully reproduces all the $\Upsilon$ measurements. This provides strong evidence that inelastic scatterings with thermal partons, rather than color screening, dominate quarkonium suppression in heavy-ion collisions.
\end{abstract}
\date{\today}
    
\maketitle
\emph{Introudction.--}
In comparison with elementary nucleon-nucleon (pp) collisions, quarkonium yields are anomalously suppressed in heavy-ion collisions at RHIC and LHC energies~\cite{STAR:2019fge,PHENIX:2006gsi,ALICE:2023gco,STAR:2022rpk,CMS:2018zza,CMS:2023lfu,STAR:2012jzy,ALICE:2020pvw} and have long been considered as a smoking gun of quark--gluon plasma (QGP) formation~\cite{Matsui:1986dk}, see for instance the reviews~\cite{Frawley:2008kk,Brambilla:2010cs,Lansberg:2019adr,Rothkopf:2019ipj,Zhao:2020jqu,Andronic:2024oxz}. However, the mechanism of quarkonium suppression in the QGP remains an open theoretical question.
In a deconfined medium, the interaction potential between a heavy quark and its antiquark is complex~\cite{Laine:2006ns,Burnier:2014ssa,Lafferty:2019jpr,Wu:2022nbv}. The real part is modified by color screening from the surrounding quarks and gluons, leading to a reduction of the confining Cornell potential. Meanwhile, the imaginary part arises from dynamical interactions with the medium, including Landau damping and singlet-to-octet transitions, resulting in a finite thermal decay width of the quarkonium state~\cite{Brambilla:2008cx,Brambilla:2013dpa,Brambilla:2011sg}. At asymptotically high temperatures, the potential can be calculated within Hard-Thermal-Loop (HTL) resummed perturbation theory~\cite{Laine:2006ns}, whereas at temperatures relevant for heavy-ion collisions it must be determined nonperturbatively. In recent years, lattice QCD calculations based on spectral reconstruction techniques~\cite{Larsen:2019zqv,Bazavov:2023dci,Ding:2025fvo}, T-matrix analyses~\cite{Tang:2023tkm,Tang:2024dkz}, Bayesian analyses~\cite{Liu:2026uav}, and machine-learning methods~\cite{Shi:2021qri} have made significant progress in extracting the in-medium heavy-quark potential. The results are qualitatively consistent, indicating that the real part of the potential remains close to its vacuum form, while the imaginary part is substantially larger than that predicted by perturbative calculations. Phenomenological studies based on the time-dependent Schr\"odinger equation also favor a strongly binding potential accompanied by sizable thermal broadening~\cite{Wen:2022utn,Wen:2022yjx,Chen:2024iil}. These nonperturbative studies challenge the long-standing picture of quarkonium suppression dominated by color screening.

Although the results of lattice QCD support the thermal decay explanation of quarkonium suppression, the comparison with experimental data is necessary for finally determining the mechanism. The difficulty is however that, almost all the used models describing quarkonium production in heavy ion collisions incorporate color screening and thermal decay simultaneously~\cite{Grandchamp:2003uw,Grandchamp:2005yw,Zhao:2010nk,Du:2017qkv,Yan:2006ve,Liu:2010ej,Zhou:2014kka,Chen:2018kfo,Zhao:2022ggw,Yao:2020xzw,Yao:2018nmy,Islam:2020bnp,Villar:2022sbv,Song:2023zma}, making it complicated to identify which mechanism is primarily responsible for the observed suppression. Furthermore, in high-energy heavy-ion collisions, especially at the LHC energies, the abundant production of heavy quarks leads to substantial quarkonium regeneration through recombination in the medium~\cite{Braun-Munzinger:2007fth,Yan:2006ve,Zhao:2010nk}, introducing a strong competition between suppression and regeneration~\cite{Zhao:2010nk,Yan:2006ve}. To identify the mechanism behind quarkonium suppression, one therefore needs both a clean experimental system and model-independent theoretical calculations. The production of $\Upsilon$ in Au+Au collisions at RHIC energy satisfies these conditions. Owing to the small production cross section of bottom quarks at $\sqrt{s_{\rm NN}}$=200 GeV~\cite{Andronic:2015wma}, $\Upsilon$ regeneration through the recombination of uncorrelated $b$ and $\bar b$ quarks is negligible compared with primordial production~\cite{Grandchamp:2005yw,Liu:2010ej,Du:2017qkv}. The measured nuclear modification factor of $\Upsilon$ therefore provides a direct probe of the in-medium suppression mechanism. On the other hand, the lattice-QCD determinations of the quarkonium masses and thermal widths~\cite{Bazavov:2023dci,Ding:2025fvo}, which are related to the real and imaginary parts of the in-medium interaction potential respectively, are available for $\Upsilon$ states and can be directly incorporated into the theoretical calculation without introducing additional model assumptions. In this Letter, we calculate the nuclear modification factor of $\Upsilon$ by considering the color-screening and thermal-decay mechanisms independently. By comparing the calculations with experimental data, we identify the dominant dynamical mechanism responsible for $\Upsilon$ suppression.  

\emph{$\Upsilon$ transport.--}
Due to its large mass, $\Upsilon$ is unlikely to be thermalized with the medium. Its phase-space distribution $f_\Upsilon$ is governed by a transport equation. Since regeneration is negligible at RHIC energy, the collision term is solely controlled by the suppression process. The relativistic Boltzmann equation for $f_\Upsilon$ is expressed as
\begin{equation}
\left[\cosh(y-\eta)\frac{\partial}{\partial \tau}+\frac{\sinh(y-\eta)}{\tau}\frac{\partial}{\partial \eta} + {\bm v}_T \cdot \nabla_T \right] f_\Upsilon = -\Gamma f_\Upsilon,
\label{boltzmann}
\end{equation}
where we have adopted the commonly used proper time $\tau$, space-time rapidity $\eta$ and momentum rapidity $y$ instead of $t$, $z$ and $p_z$ respectively, and ${\bm v}_T = {\bm p}_T / \sqrt{m_\Upsilon^2 + p_T^2}$ is the transverse velocity of $\Upsilon$. The suppression rate $\Gamma(T)$, determined by the underlying suppression mechanism, depends on the medium temperature $T$.  

The local temperature  $T({\bm x}_T, \eta, \tau)$ of the QGP is extracted from relativistic hydrodynamic simulations coupled to an appropriate equation of state (EoS). In this Letter, we employ the MUSIC package~\cite{Schenke:2010nt,McDonald:2016vlt} to simulate the evolution of the hot medium and adopt a lattice-QCD-based EoS featuring a smooth crossover between the QGP and the hadron resonance gas at the critical temperature $T_C\approx160~\rm MeV$~\cite{HotQCD:2014kol}. For the most central collisions at central rapidity, the maximum temperature of the medium is chosen to be $T_{max}({\bm x}_T = 0,\eta=0,\tau_0)$ = 391 MeV at the initial time $\tau_0=0.6\ fm/c$, obtained by fitting the experimentally measured charged-particle multiplicity in Au+Au collisions at RHIC energy.

The transport equation (\ref{boltzmann}) can be solved analytically. In the central rapidity region with $\eta,\ y\sim 0$, the solution reads
\begin{eqnarray}
\label{solution} 
    f_\Upsilon({\bm x}_T, {\bm p}_T, \tau) &=& f_{ini}\left({\bm x}_T(\tau,\tau_0),{\bm p}_T\right)\nonumber \\
    && \times e^{ -\int_{\tau_0}^\tau \Gamma(T({\bm x}_T(\tau, \tau'), \tau'))d\tau'},   
\end{eqnarray}
where the leakage effect in the transverse direction modifies the transverse coordinates~\cite{Yan:2006ve} ${\bm x}_T(\tau,\tau_0)={\bm x}_T-{\bm v}_T(\tau-\tau_0)$ and ${\bm x}_T(\tau,\tau')={\bm x}_T-{\bm v}_T(\tau-\tau')$. Integrating the distribution function over phase space, one obtains the $\Upsilon$ yield $N_{AA}(\Upsilon)$ in heavy ion collisions.

The analytical solution (\ref{solution}) depends on the initial $\Upsilon$ distribution $f_{ini}({\bm x}_T,{\bm p}_T)$. Since we focus on the central rapidity region, the initial transverse distribution can be factorized into spatial and momentum components.
The spatial component is governed by the number of binary pp collisions. Within the Glauber model~\cite{Miller:2007ri}, the spatial density for a given impact parameter ${\bm b}$ is proportional to the binary collision density,
\begin{equation}
\frac{d^2 N_{ini}}{d^2{\bm x}_T} = \sigma_{pp} \, T_A\left({\bm x}_T + 
       {\bm b}/2\right)T_B\left({\bm x}_T - 
       {\bm b}/2\right),
\label{spatialinit}
\end{equation}
where $\sigma_{pp}=42~\mathrm{mb}$~\cite{Miller:2007ri} 
is the inelastic pp cross section at $\sqrt{s_{\rm NN}}=200~\mathrm{GeV}$, and $T_{A,B}$ are the thickness functions of the colliding nuclei $A$ and $B$. The impact parameter ${\bm b}$ is related to the number of participant nucleons $ N_{part}$, which characterizes the collision centrality. The initial momentum distribution in heavy-ion collisions is obtained by superposing the corresponding pp distributions~\cite{Chen:2018kfo,CDF:2001fdy}.

We now briefly estimate the cold nuclear matter effect on $\Upsilon$ production. Owing to the large bottom quark mass and the compact size of the $\Upsilon$, nuclear absorption is expected to be significantly weaker than that for charmonium. As for nuclear shadowing, a simple estimate of the longitudinal momentum fraction carried by the two initial gluons in the fusion process $gg\to\Upsilon$ gives $x\sim M_\Upsilon/\sqrt{s_{\rm NN}}\approx 0.05$ at RHIC energy. This corresponds to the transition region between shadowing and antishadowing, where the nuclear modification factor of the gluon distribution is close to unity. Therefore, cold nuclear matter effects can be safely neglected in the present study. This approximation is further supported by d+Au measurements~\cite{STAR:2013kwk} and other theoretical studies~\cite{Ferreiro:2011xy,Rakotozafindrabe:2012ss}. 
  
Note that each $\Upsilon$ state $i$ ($i=1S,\ 1P,\ 2S$, $2P,\ 3S$) satisfies its own transport equation (\ref{boltzmann}). After the evolution in the hot medium, the final yield of a given state contains both the direct production and the feed-down contributions from excited states~\cite{ParticleDataGroup:2022pth}. Taking $\Upsilon(1S)$ as an example, the inclusive nuclear modification factor $R_{AA}(1S)$ can be expressed in terms of the direct nuclear modification factors $R_{AA}^{dir}(i)$, 
\begin{equation}
R_{AA}(1S) = \frac{N(1S)}{N_{pp}(1S) N_{coll}} = \sum_i R_{AA}^{dir}(i) f_i,
\label{RAA_inclusive}
\end{equation}
where $N$ and $N_{pp}$ are the inclusive $\Upsilon(1S)$ yield in $AA$ and $pp$ collisions, $N_{coll}$ is the number of binary pp collisions, controlled by the nuclear geometry, and $f_i$ represents the feed-down contribution from state $i$ to the inclusive $\Upsilon(1S)$ yield, determined from experimental pp data~\cite{Islam:2020bnp,Wen:2022yjx,Lansberg:2019adr,CMS:2018zza}.  

\emph{Color screening.--}
Let us first consider the effect of color screening on the properties of heavy quarkonia. When a heavy quarkonium is placed in a dense QGP, the surrounding colored quarks and gluons gradually screen the color interaction between the heavy quark and antiquark. When the screening length $\lambda_D$ becomes smaller than the distance between the two heavy quarks, the bound state dissociates in the medium. This is the color-screening picture, analogous to electric screening in electrodynamics. The screening length $\lambda_D$, or equivalently the screening mass $m_D = 1/\lambda_D$, can be calculated perturbatively within finite-temperature QCD~\cite{Laine:2006ns,Brambilla:2008cx,Brambilla:2013dpa,Brambilla:2011sg} or extracted from lattice-QCD calculations~\cite{Burnier:2014ssa,Lafferty:2019jpr}.

Color screening leads to the sudden dissociation of quarkonia once the temperature exceeds the screening or melting temperature $T_D$. The suppression rate is therefore given by
\begin{equation}
	\Gamma(T) = 
	\begin{cases} 
		\infty & \text{if } T > T_D, \\
		0      & \text{if } T \le T_D,
	\end{cases}
	\label{td}
\end{equation}
and the transport solution (\ref{solution}) becomes
\begin{equation}
	f_\Upsilon({\bm x}_T, {\bm p}_T, \tau) = f_{\text{ini}}({\bm x}_T(\tau,\tau_0), {\bm p}_T)\Theta(T_D - T),
	\label{f_screening}
\end{equation}
where $\Theta$ is the step function describing the sudden melting.

\begin{figure}[htbp]
    \centering
    \includegraphics[width=0.95\linewidth]{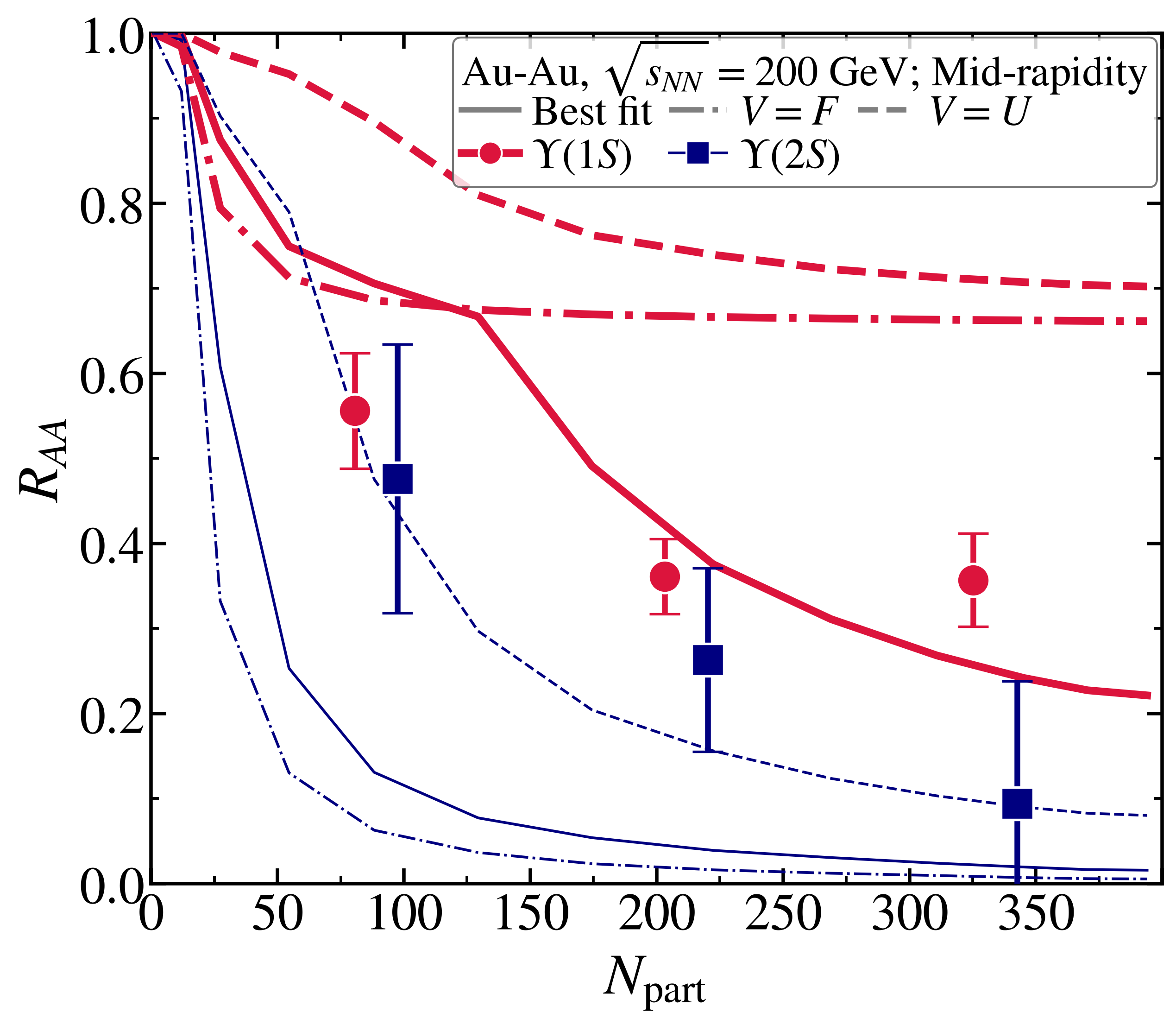}\\
    \includegraphics[width=0.95\linewidth]{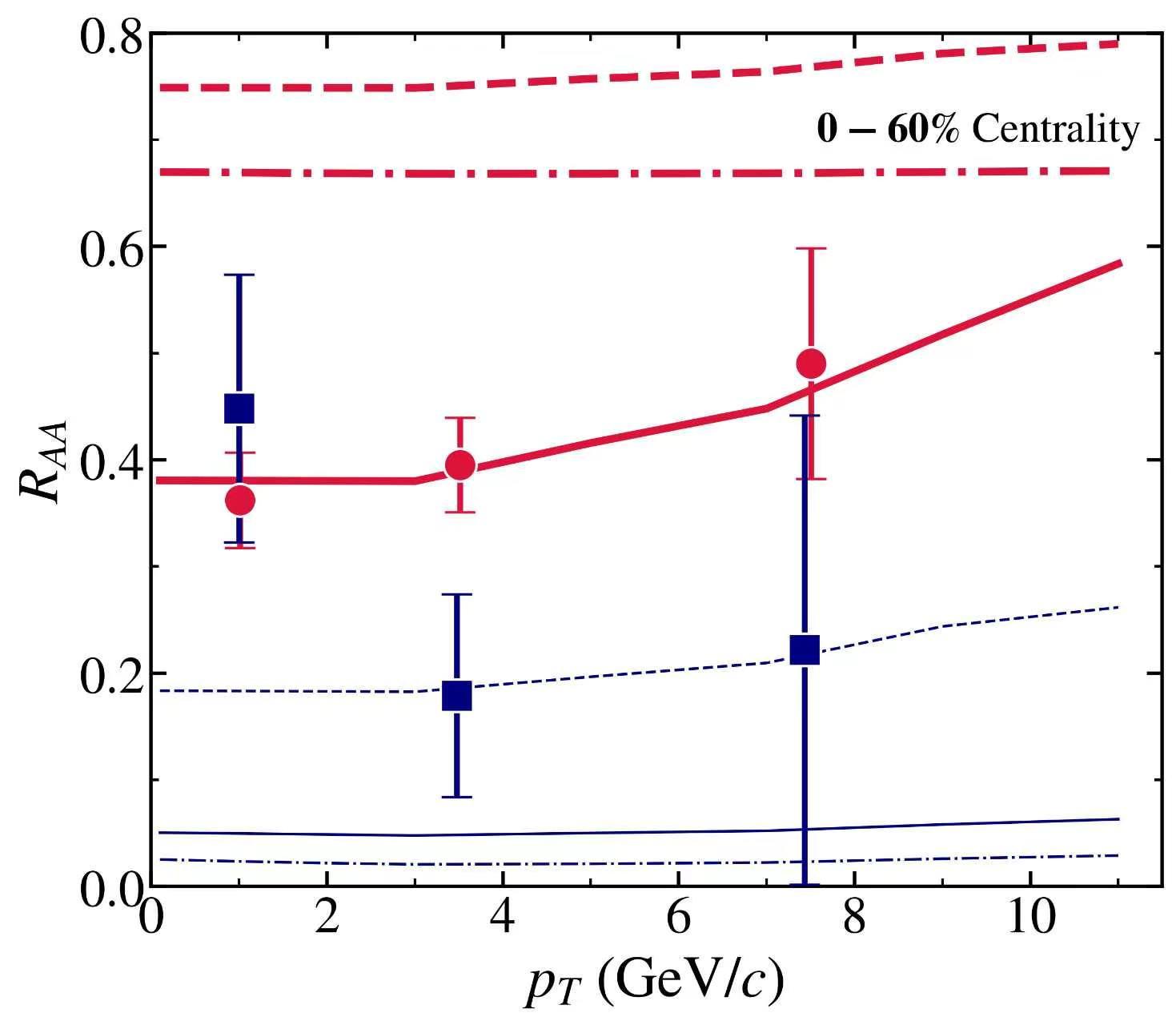}
    \caption{Nuclear modification factor $R_{AA}$ as a function of $N_{part}$ (upper panel) and $p_T$ (lower panel) for $\Upsilon(1S)$ (thick lines) and $\Upsilon(2S)$ (thin lines) in Au+Au collisions at $\sqrt{s_{\rm NN}} = 200$ GeV. The dashed, dot-dashed, and solid lines are the calculations with the melting temperatures $T_D$ extracted from the internal energy $U$, free energy $F$ and best fit respectively. The experimental data are taken from Ref.~\cite{STAR:2022rpk}.}
    \label{fig1}
\end{figure}

The quantity directly calculated in lattice QCD is the heavy-quark free energy $F(T)$. If the heavy-quark potential is taken as $V=F$ and inserted into the two-body Schr\"odinger equation for the $b\bar b$ system, the temperature at which the binding energy satisfies $\epsilon(T)=0$ is defined as the melting temperature $T_D$~\cite{Satz:2005hx,Zhao:2020jqu}. For the $\Upsilon$ states $1S, 1P, 2S, 2P$ and $3S$, the scaled melting temperature $T_D/T_C$ is $3.00, 1.12, 1.08, 1.00$ and $<1$, respectively. If the internal energy $U = F - T\partial F/\partial T$ is instead adopted as the heavy-quark potential, the Schr\"odinger equation gives the corresponding melting temperatures $>4.00, 1.76, 1.60, 1.19$ and $1.17$. Taking these melting temperatures in Eq.~(\ref{f_screening}), the resulting centrality and transverse-momentum ($p_T$) dependence of the nuclear modification factor $R_{AA}$ for the states $1S$ and $2S$ are shown in Fig.~\ref{fig1}. While the choice of $V=U$ can well describe the centrality and $p_T$ dependence of $\Upsilon(2S)$ reasonably well, the remaining cases deviate significantly from the experimental data. This indicates that neither $U$ nor $F$ can simultaneously describe the suppression of $\Upsilon(1S)$ and $\Upsilon(2S)$. 

It is natural to ask whether the melting temperature obtained from the nonrelativistic Schr\"odinger equation depends sensitively on the choice of the heavy-quark potential. To examine whether the color-screening picture can account for the experimental data, we determine the melting temperatures by directly fitting the data using the Nelder--Mead simplex optimization algorithm~\cite{Nelder:1965zz}. The best fit to $R_{AA}(N_{part})$, corresponding to the minimum of the $\chi^2$ function,
\begin{equation}
	\chi^2=\sum_n{\left(R_{AA}^{th,n}-R_{AA}^{exp,n}\right)^2\over \left(\delta R_{AA}^{exp,n}\right)^2}
\end{equation}
with the theoretical and experimental nuclear modification factors $R_{AA}^{th,n}$ and $R_{AA}^{exp,n}$ at the $n$-th data point and the combined experimental uncertainty $\delta R_{AA}^{exp,n}$, is shown as solid lines in Fig.~\ref{fig1}. The extracted melting temperatures are $T_D/T_C = 1.97, 1.19, 1.19, 1.12, 1.01$. While for $\Upsilon(1S)$ the $p_T$ dependence is very good and the centrality dependence is qualitatively good, the description for $\Upsilon(2S)$ is far from the data. Note that only $R_{AA}(N_{part})$ is included in the fit, whereas $R_{AA}(p_T)$ is a prediction of the model.
We therefore conclude that the color-screening mechanism alone cannot simultaneously describe the suppression of the ground and excited $\Upsilon$ states.

\emph{Thermal decay.--}
We now turn to the description of quarkonium suppression within the thermal-decay picture. In contrast to the sudden melting induced by color screening, the thermal decay of the $\Upsilon$ is driven by inelastic scattering with thermal partons and therefore proceeds through a continuous suppression rate throughout the QGP evolution. The $\Upsilon$ suppression caused by thermal decay can be equivalently described by introducing a complex interaction potential in quantum mechanics~\cite{Rothkopf:2019ipj}.
\begin{figure}[htbp]
	\centering
	\includegraphics[width=0.95\linewidth]{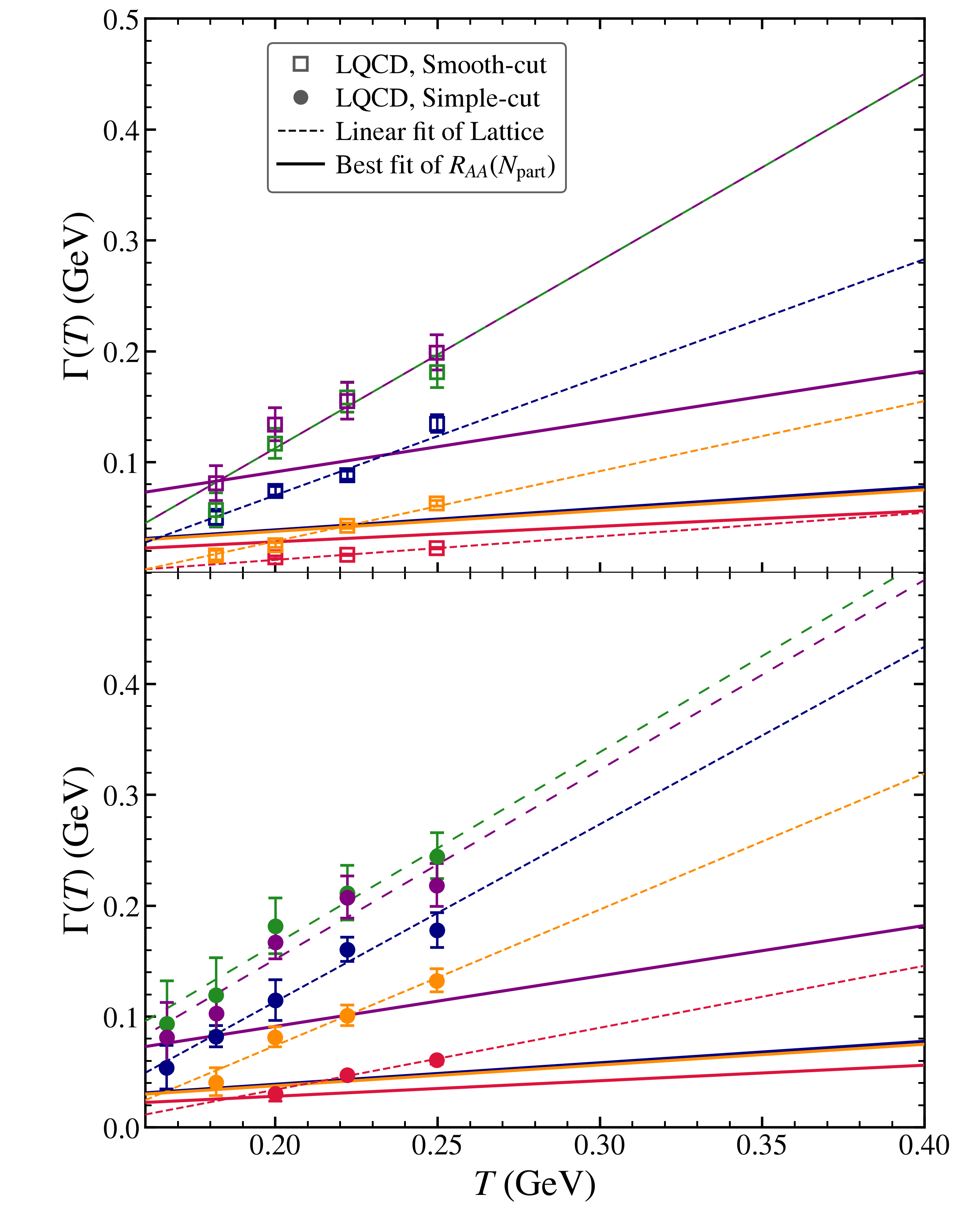}
	\caption{Thermal decay width as a function of $T$. The open squares and solid circles are lattice QCD results obtained using the smooth-cut~\cite{Ding:2025fvo} (upper panel) and simple-cut~\cite{Bazavov:2023dci} (lower panel) Lorentzian parametrizations, respectively. The dashed lines are the linear fit of the lattice data, and solid lines are from the best fit of the experimentally measured $R_{AA}(N_{part})$, shown in Fig.~\ref{fig3}. The red, yellow, blue, purple and green symbols and lines correspond to the $1S,\ 1P,\ 2S,\ 2P$ and $3S$ states, respectively. }
	\label{fig2}
\end{figure}

\begin{figure}[htbp]
	\centering
	\includegraphics[width=0.95\linewidth]{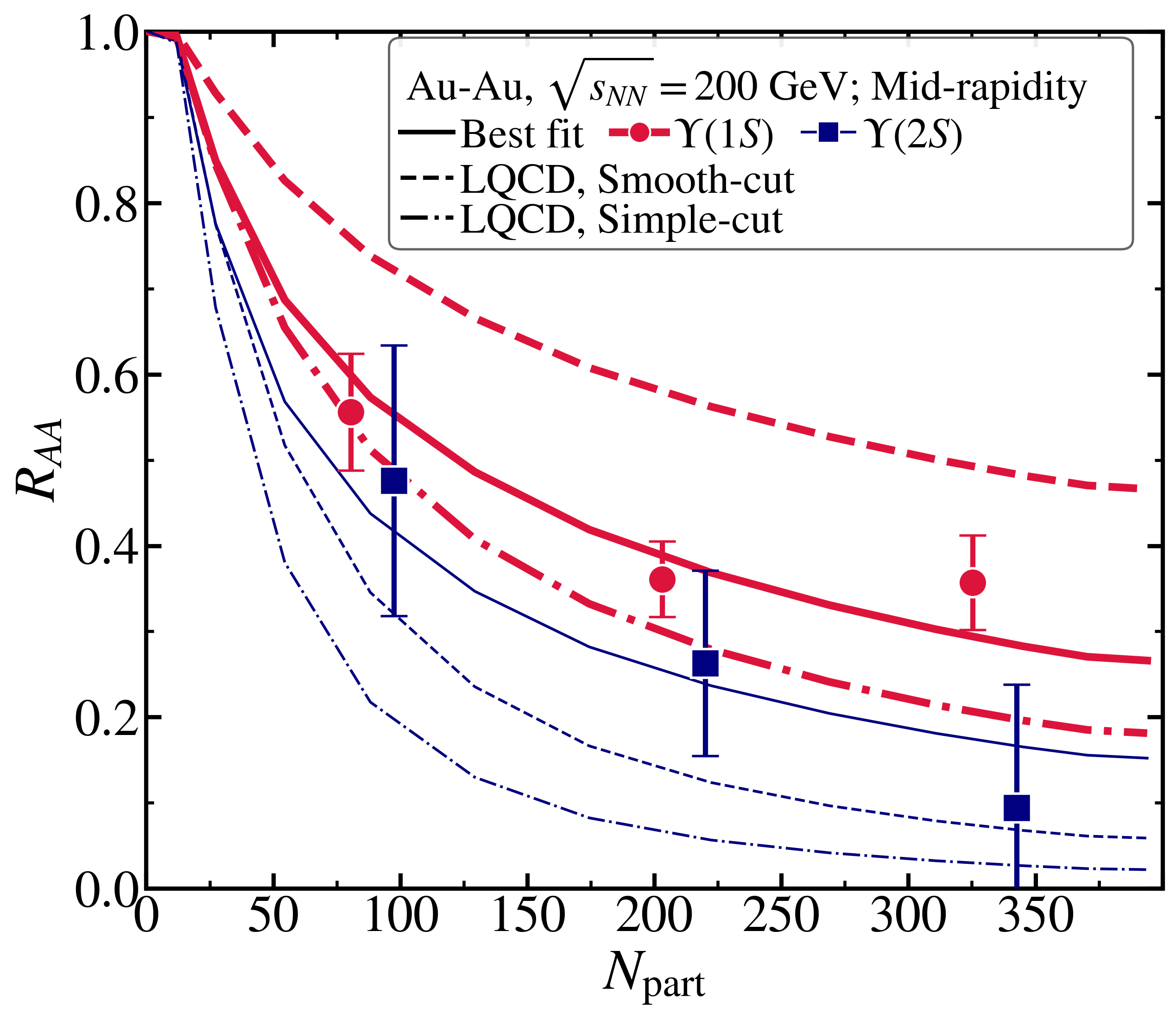}\\
	\includegraphics[width=0.95\linewidth]{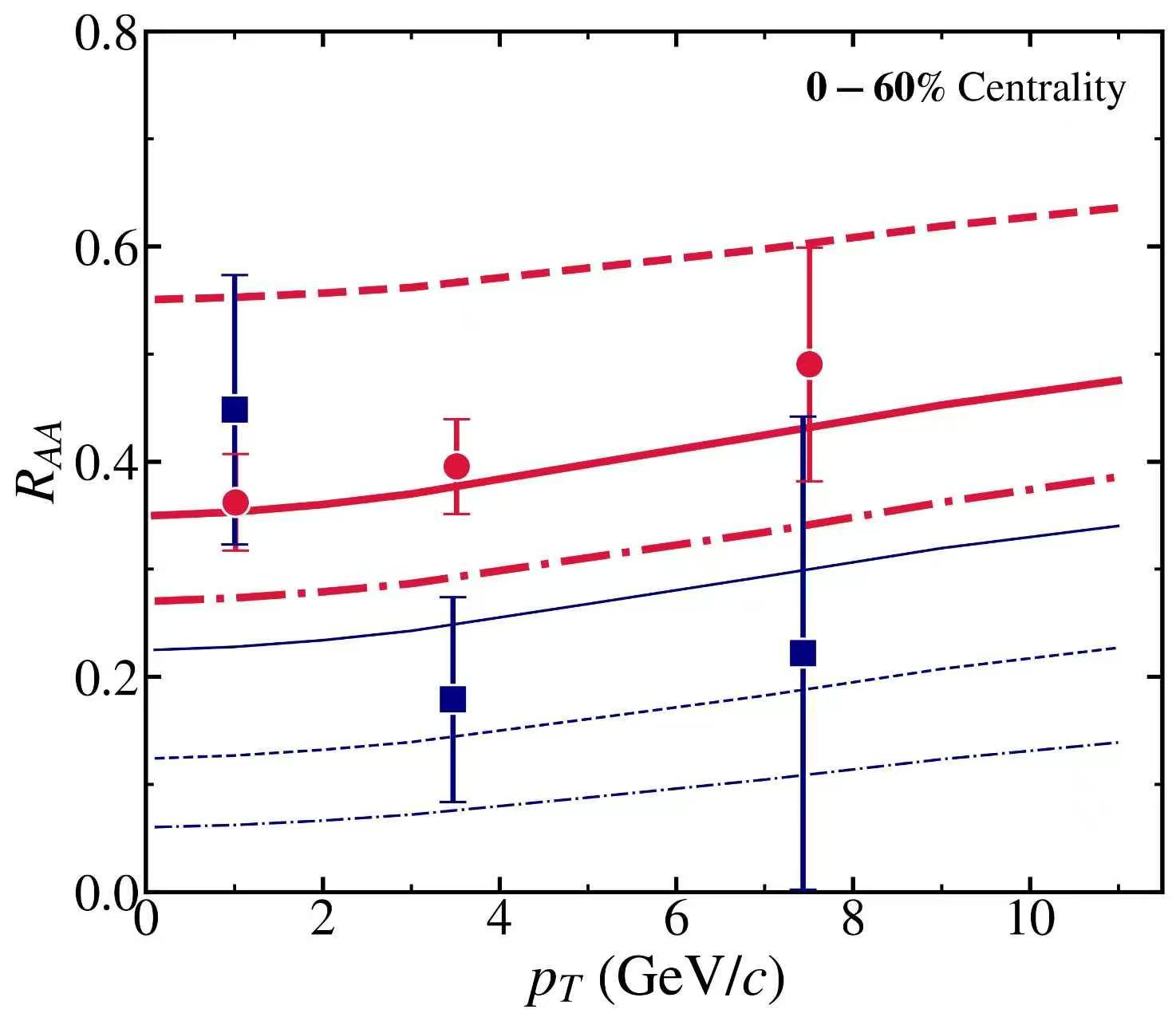}
	\caption{Nuclear modification factor $R_{AA}$ as a function of $N_{\text{part}}$ (upper panel) and $p_T$ (lower panel) for $\Upsilon(1S)$ (thick lines) and $\Upsilon(2S)$ (thin lines) in Au+Au collisions at $\sqrt{s_{\rm NN}} = 200$ GeV. The dashed and dot-dashed lines show the calculations using the smooth-cut and simple-cut lattice widths, the solid lines are the best fit by taking the linear widths shown as solid lines in Fig.~\ref{fig2}, and the experimental data are taken from Ref.~\cite{STAR:2022rpk}.}
	\label{fig3}
\end{figure}

To reduce model dependence, instead of using a parameterized imaginary potential, we directly incorporate the lattice-QCD decay widths $\Gamma_i(T)$ into the transport solution (\ref{solution}) for the $\Upsilon$ computation. Fig.~\ref{fig2} shows the lattice results obtained using the smooth-cut~\cite{Ding:2025fvo} and simple-cut~\cite{Bazavov:2023dci} Lorentzian parametrization. 
For each state $i$, the temperature dependence of the decay width is well approximated by a linear parametrization, shown as dashed lines in Fig.~\ref{fig2}. In the upper panel, the curves for the $2P$ and $3S$ states almost coincide. Using these lattice-QCD decay widths directly, the calculated $R_{AA}$ is shown as dashed (smooth-cut) and dot-dashed (simple-cut) lines in Fig.~\ref{fig3}. From the comparison with the experimental data, the two groups of lattice widths can both describe reasonably well the trend of $R_{AA}(N_{part})$ and $R_{AA}(p_T)$ for the ground and excited states.       

Considering that the lattice QCD results depend on the parameterization of the spectral function, as shown in the upper and lower panel of Fig.~\ref{fig2}, and still suffer from sizeable uncertainties in the calculations, the thermal decay widths cannot yet be determined precisely from lattice QCD. To explore whether the observed $\Upsilon$ suppression can be quantitatively explained by an appropriate thermal decay width, we adopt a linear parametrization, $\Gamma_i(T)=a_i T$, and determine the slopes $a_i$ by fitting the experimental data $R_{AA}(N_{part})$. The best fit for the states $1S$ and $2S$ are shown as thick and thin solid lines in Fig.\ref{fig3}, and the corresponding thermal decay widths $\Gamma_i$ are shown as solid lines in Fig.~\ref{fig2}. Since the feed-down contribution from $3S$ to $1S$ is only $0.9\%$, $R_{AA}(1S)$ and $R_{AA}(2S)$ are not sensitive to the $3S$ yield and cannot constrain the width $\Gamma_{3S}$. This is the reason why we did not show $\Gamma_{3S}$ in Fig.~\ref{fig2}. Compared with the RHIC data, the best fit provides a good description of both the centrality and $p_T$ dependence of $\Upsilon(1S)$ and $\Upsilon(2S)$. The extracted thermal widths are reasonable: they are larger than the perturbative calculations and comparable to the lattice-QCD results.

\emph{Summary and discussion.--}
We investigated $\Upsilon$ production in Au+Au collisions at $\sqrt{s_{\rm NN}}=200$\ GeV, aiming to clarify the mechanism behind quarkonium suppression in high energy nuclear collisions. Since the $\Upsilon$ regeneration in the hot medium can be safely neglected, the $\Upsilon$ phase space distribution is solely governed by the suppression rate, making it easier to distinguish different suppression mechanisms. By comparing our calculations with the RHIC data, we found that no color-screening temperature can simultaneously explain the suppression of both the ground and excited $\Upsilon$ states, whereas an appropriate thermal decay width provides a good description of the suppression of all measured $\Upsilon$ states.

The failure of the color-screening mechanism originates from its abrupt suppression picture: all $\Upsilon$s above the screening temperature $T_D$ are completely dissociated, while all below $T_D$ survive. Suppose the QGP is a uniform fireball with temperature satisfying $T_C < T < T_{max}$, and consider only one excited state for simplicity, say $\Upsilon(2S)$. For the case of $T_D(1S) > T_D(2S) > T_{max}$, there is no suppression for any $\Upsilon$ state,  $R_{AA}(1S) = R_{AA}(2S) = 1$; For the case of $T_{max} > T_D(1S) > T_D(2S)$, all the $\Upsilon$ states are eaten up by the medium, $R_{AA}(1S) = R_{AA}(2S) = 0$; For the case of $T_D(1S) > T_{mnax} > T_D(2S)$, the excited state disappears but the ground state survives, $R_{AA}(1S) \sim 0.5,\ R_{AA}(2S) = 0$. Therefore, any combination of $T_D(1S)$ and $T_D(2S)$ with $T_D(1S) > T_D(2S)$ cannot explain the experimental observation that both $R_{AA}(1S)$ and $R_{AA}(2S)$ are clearly nonzero. Note that the sudden turns in the lines in Fig.~\ref{fig1} arise from the sudden dissociation of different $\Upsilon$ states. In contrast, thermal decay induced by inelastic scatterings can occur at any temperature throughout the QGP evolution. Consequently, $\Upsilon$ suppression takes place continuously at every stage of the fireball evolution, allowing both the ground and excited $\Upsilon$ states to survive with finite probabilities.

At the high temperature of $2T_C$, the lattice QCD result for $\Upsilon$~\cite{Larsen:2019zqv,Shi:2021qri,Ding:2025fvo} shows still temperature independent mass and linearly increasing width. If this trend holds at higher temperatures, the conclusion that the $\Upsilon$ suppression is dominated by thermal decay will be still true in the LHC energy region. However, the regeneration becomes important in this case, the competition between suppression and regeneration will control the $\Upsilon$ production in heavy ion collisions at LHC. 

\vspace{0.3cm}
\noindent {\bf Acknowledgment}: 
We acknowledge the discussions with Dr. Shile Chen in the beginning of this work. The work is supported by Yantai university under the grant 2226001 and the NSFC Grant 12575149.  
\vspace{1cm}

\bibliographystyle{apsrev4-2}
\bibliography{Ref}
\end{document}